\documentclass{cambridge6A}
  \usepackage{graphicx}
  \usepackage{amssymb,amsmath,amsfonts}

\begin{document}


\newcommand \be{\begin{equation}}
\newcommand \ee{\end{equation}}
\newcommand \bea{\begin{eqnarray}}
\newcommand \eea{\end{eqnarray}}
\newcommand \bee{\begin{equation}}
\newcommand \eee{\end{equation}}

\def\half{{1\over 2}}
\def\nref#1{(\ref{#1})}

\renewcommand*\thesection{\arabic{section}}
\renewcommand*\thesubsection{\arabic{section}.\arabic{subsection}}
\setcounter{equation}{0}
\numberwithin{equation}{section}
\setcounter{figure}{0}
\renewcommand{\thefigure}{\arabic{figure}.}




\chapterauthor{Juan Maldacena
\\ ~ \\
\textit{ Institute for Advanced Study, Princeton, NJ 08540, USA}
}

\chapter*{The gauge/gravity duality}

\contributor{Juan Maldacena
\affiliation{Institute for Advanced Study }
}

\begin{abstract}\small
Short introduction to the gauge gravity duality.
\end{abstract}

\copyrightline{Chapter of the book \textit{Black Holes in Higher Dimensions} to
be published by Cambridge University Press (editor: G. Horowitz)}

In this chapter we explain the gauge/gravity duality \cite{JM,GKP,wittenhol}
which is one of the motivations for studying black
hole solutions in various numbers of dimensions.
The gauge/gravity duality is an equality between two theories: On one side we have a quantum field
theory in $d$ spacetime dimensions. On the other side  we have a gravity theory on a $d+1$
 dimensional spacetime that has
an asymptotic boundary which is $d$ dimensional.
It is also sometimes called $AdS/CFT$, because the simplest examples involve anti-de-Sitter spaces and
conformal field theories. It is often called   gauge-string duality. This is
 because the gravity theories
are  string theories and the quantum field theories are gauge theories. It is also referred to as
``holography'' because one is describing a $d+1$ dimensional gravity theory in terms of a lower
dimensional system, in a way that is reminiscent of an optical hologram which stores a three dimensional
imagine on a two dimensional photographic plate. It is called a ``conjecture'', but by now there is a
lot of evidence that it is correct. In addition, there are some
derivations based on physical arguments.

The simplest example involves an anti-de-Sitter spacetime. So, let us start describing this
spacetime in some detail. Anti-de-Sitter is the simplest solution of Einstein's equations with
a negative cosmological constant. It is the lorentzian analog of hyperbolic space, which was
historically the first example of a non-Euclidean geometry. In a similar way,
AdS/CFT gives the simplest example of a
quantum mechanical spacetime.

\begin{figure}
\begin{center}
  \includegraphics[height=2.in]{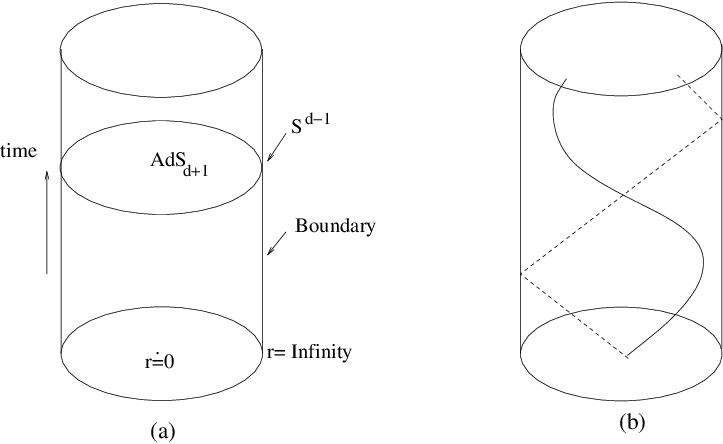}\\
  \caption{(a) Penrose diagram for Anti-de-Sitter space. It is a solid cylinder.
  The vertical direction is time. The boundary contains the time direction and a sphere, $S^{d-1}$, represented here as a circle. (b) Massive geodesic (solid line) and a massless geodesic (dashed line). }\label{Penrose}
  \end{center}
\end{figure}

The metric in $AdS$ space can be written as
\be \label{adsmet}
ds^2_{AdS_{d+1}}  = L^2 \left[  -(r^2 +1 ) dt^2 + { dr^2 \over r^2 +1 } + r^2 d\Omega_{d-1}^2 \right]
\ee
where the last term is the metric of a unit sphere, $S^{d-1}$. $L$ is the radius of curvature.
Note that near $r=0$ it looks like flat space.
As we go to larger values of $r$ we see that $g_{00}$ and the metric on the sphere grow.
The growth of $g_{00}$ can be viewed as a rising gravitational potential. In fact, a slowly moving
massive particle feels a gravitational potential $V \sim \sqrt{-g_{00}}$. If a particle is set at
rest at a large value of $r$, it will execute an oscillatory motion in the $r$ direction, very much
like a particle in a harmonic oscillator potential.
This gravitational potential confines particles around the origin. A massive particle with
finite energy cannot escape to infinity, $r = \infty$.
A massless geodesic can go to infinity and back in
finite time. One way to see this is to look at the Penrose diagram of $AdS$. We can factor out a
factor of $1+r^2$ in the metric \nref{adsmet}, and define a new radial coordinate, $x$, via
$dx = dr/(1+r^2)$ which now has a finite range. Thus, the Penrose diagram of $AdS$ space is
a solid cylinder, see figure \ref{Penrose}(a).
 The vertical direction is time, the boundary is at $r = \infty$, which is a
finite value of $x$. The $S^{d-1}$ is the spatial section of the surface of the cylinder.
The metric in \nref{adsmet} has an obvious $R\times SO(d)$ symmetry. $AdS$ has more
symmetries. The full symmetry group is $SO(2,d)$.
 These symmetries can be made more manifest by viewing $AdS$ as the
hyperboloid
\be
 -Y_{-1}^2 - Y_0^2 + Y_1^2 + \cdots Y_{d}^2 = - L^2
 \ee
  in $R^{2,d}$. This description is
useful for realizing the symmetries explicitly. However, in this hyperboloid, the time direction, $t$
in \nref{adsmet}, is compact (it is
just the angle in the $[-1,0]$ plane). However, in all physical applications
we want to take this time direction to be non-compact.

These isometries of $AdS$ are very powerful. Let us recall the situation in flat space. If we have
a massive geodesic in flat space we can always boost to a frame where is at rest. In $AdS$ it is the same,
if we consider the oscillating trajectory of a massive particle, then we can ``boost'' to a frame
where that particle is at rest. Thus, the moving particle does not know that it is moving and, despite
appearances, there is no ``center'' in $AdS$. The Hamiltonian is part of the symmetry group (as in
the Poincare group) and there are several choices for a Hamiltonian. Once we choose one Hamiltonian,
for example the one that shifts $t$ in \nref{adsmet}, then we choose a ``center'' and a notion of
lowest energy state, which is  a particle sitting at this ``center''.

In some applications it is useful to focus
on a small patch of the boundary and treat it as $R^{1,d}$. In fact, there is a choice of coordinates
where the $AdS$ metric takes the form
\be \label{adsmep}
ds^2 = L^2 {  - dt^2 + d \vec x^{\, 2}_{d-1} + dz^2 \over z^2 }
\ee
Here the boundary is at $z=0$ and we have slices that display the Poincare symmetry group in $d$ dimensions (one time and $d-1$ spatial dimensions). In fact, if we take $t \to i x_0$ we get hyperbolic space, now
sometimes called Euclidean AdS!. In these coordinates we can also see clearly another isometry which
rescales the coordinates $ (t,\vec x , z) \to \lambda ( t, \vec x,z)$. These coordinates have a horizon
at $z=\infty$ and cover only a portion of \nref{adsmet}. These coordinates are
 convenient when we want do consider a CFT on living in Minkowski space, $R^{1,d-1}$.

The $AdS/CFT$ relation postulates that all the physics in an asymptotically anti-de-Sitter
spacetime  can be described by a local quantum
field theory that lives on the boundary. The boundary is given by $R \times S^{d-1}$. The isometries of
$AdS$ act on the boundary. They send points on the boundary to points on the boundary. This action
is simply the action of the conformal group in $d$ dimensions, $SO(2,d)$. Thus, the quantum field
theory is a conformal field theory.  In fact,  the rescaling symmetry of
 \nref{adsmep} translates into a dilatation on the boundary. The boundary theory is thus scale
 invariant. It has no dimensionfull parameter. Usually theories that are scale invariant are also
 conformal invariant. These are theories where the stress energy momentum tensor is traceless.
  The conformal group includes the poincare group, the dilatation, and ``special conformal transformations'', which will not be too  important for us here. Notice that the conformal symmetry makes
 sure that we can choose an arbitrary radius for the boundary $S^{d-1}$, so we can set it to one.
 In fact, if we have a conformal field theory, the tracelessness of the stress tensor implies
 that a field theory on a space with a metric $g^b_{\mu \nu} $ or $\omega^2(x) g^b_{\mu\nu}$ is basically
 the same (up to a well understood conformal anomaly). Here we are talking about the metric on the
 boundary,  where the field theory lives. This metric is not dynamical, it is fixed.

 How can it be that a $d+1$ dimensional bulk theory is equivalent to a $d$ dimensional one?
 In fact, let us attempt to disprove it. A skeptic would argue as follows. Let us do a simple count
 of the number of degrees of freedom. Since the bulk has one extra dimension, we seem to  have
 a contradiction. In fact, we could consider the number of degrees of freedom at large energies, in
 the microcanonical ensemble. To compute it we can introduce an effective  temperature.
   In a theory with
 massless fields (or a theory with no scale) we expect that the entropy should go like $S \sim {
 V_{d-1}T ^{d-1}}$. So if the boundary theory is a CFT on $R \times S^3$, then for large
 temperatures compared to the radius of $S^3$ ($T \gg 1$),
  we expect that the entropy should grow like
 \be \label{entropyft}
  S \propto   c \,  { T^{ d-1} }
  \ee
  where $c$  a dimensionless
    constant that measures the effective number of fields in the theory.
For free fields it can be explicitly computed, as we will do later in an example.
On the other hand, from the bulk point of view it seems that we also have a theory with
massless particles, which are the gravitons. We could also have extra fields, but, for the time being, let
us include only the gravitons,  which  give a lower bound to the entropy. The entropy of these gravitons is certainly bigger than the entropy from the region where $r\sim 1$. In that region which has a volume
of order one, we get
\be \label{snaive}
S_{\rm gas~of~gravitons }  > { T^{d} }
\ee
because it has $d$ spatial dimensions.
For large enough $T$ we see that \nref{snaive} is bigger than \nref{entropyft}.
So we appear to have a contradiction with the basic claim of $AdS/CFT$. However, we are forgetting
something essential: It is crucial that the bulk theory contains {\it gravity}.
And gravity gives rise to black holes. And black holes give rise to bounds on entropy.
Black holes in $AdS$ have the form
\be \label{adsbhmet}
ds^2_{AdS_{d+1}}  = L^2 \left[  -(r^2 +1 - { 2 g m \over r^{d-2}}  ) dt^2 + { dr^2 \over r^2 +1 - { 2 g m \over r^{d-2}}} + r^2 d\Omega_{d-1}^2 \right]
\ee
 where  $m$ is proportional to  the mass and
$g$ is proportional to   the Newton constant in units of the $AdS$ radius
\be
\label{geff}
g  \propto  { G^{d+1}_N \over L^{d-1} }
\ee
 The gas of gravitons extends up  to $r_z \sim T$ and has a mass
 of the order of $m \sim T^{d+1}$. For large $T$ we can neglect the 1 in \nref{adsbhmet} when
 computing the Schwarschild radius:  $r_s^{d} \sim g m \sim g T^{d+1}$.
 We see that the Schwarschild radius is bigger than the size of the system for temperatures that are
 big enough,
   $ T> 1/g$. Thus, the computation in \nref{snaive} breaks down for such large temperatures.
   At large enough energies we compute the entropy in terms of the black hole entropy. This
   entropy grows like the area of the horizon $ S \sim { r_{s}^{d-1} \over g} $.
 One can
    see that the Hawing temperature for big black holes is $ T \propto r_s$.
   The entropy of the black hole is $S_{BH} \sim { 1 \over g } {T^{d-1} } $.
   This is now  of the expected form, \nref{entropyft}, with
    \be \label{centralexpr}
    c  \propto { 1 \over g } \propto  {L_{AdS}^{d-1} \over G_{N,d+1}}
    \ee
Thus,
 AdS/CFT connects the entropy of a black hole with the ordinary thermal entropy of a field theory. This has two very important applications. First, for conceptual issues about the entropy of black holes,
it gives a statistical interpretation for black hole entropy. In addition, since it displays the
black hole as an ordinary thermal state in a unitary quantum field theory, we see that these
black holes are consistent with quantum mechanics and
unitary evolution. Second, it allows us to compute the thermal
free energy, and other thermal properties, in quantum field theories that have gravity duals.

 The number of fields scales like the inverse Newton constant. Notice that $g$, \nref{centralexpr},
 measures the
 effective gravitational coupling at the $AdS$ scale. It is the dimensionless constant measuring
 the effective non-linear interactions among gravitons. Thus, {\it  if we want a weakly coupled bulk theory
 we need that the field theory has a large number of fields.}. This is a necessary, but not sufficient,
 condition.
\begin{figure}
\begin{center}
  \includegraphics[height=1.in]{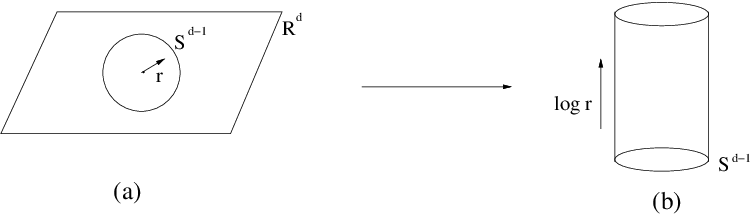}\\
  \caption{(a) A conformal field theory on the Euclidean plane, $R^d$. We can act with various operators
   at $r=0$. These create certain states on $S^{d-1}$ which are given by performing the path integral
   of the field theory in the interior of the $S^{d-1}$ with the operators inserted.
   (b) Due to the Weyl symmetry of the theory we can rescale the metric and view it as the metric
   on a cylinder $R \times S^{d-1}$. States of the theory on this cylinder are the same are in one
   to one correspondence to operators on the plane. This is a general property of CFT's and completely
   independent of $AdS/CFT$.  }\label{planetocylinder}
  \end{center}
\end{figure}
 One important feature of a weakly coupled theory is the existence of a Fock space structure in
 the Hilbert space. Namely, we can talk about a single particle, two particles, etc. Their energies
 are, up to small corrections, proportional to the sum of the energies of each of the particles.
 The dual quantum field theory has to have a similar structure. In fact, this structure emerges
 quite naturally in large $N$ gauge theories. A large $N$ gauge theory is a gauge theory based
 on the gauge group $SU(N)$ (or $U(N)$) with fields in the adjoint representation. In this case
 we can form gauge invariant operators by taking traces of the fundamental fields, such as
 $Tr[ F_{\mu \nu} F^{\mu \nu } ]$ or $Tr[ F_{\mu \nu } D_{\rho } D_{\sigma} F^{\mu\nu } ] $, etc.
 These are all local operators where the fields are all evaluated at the same point in spacetime.
 In addition, one could have double trace operators, such as the product of the two operators we
 mentioned above. When we act with one such operator on the the field theory vacuum we create
 a state in the field theory.
  In a general CFT (even if it does not have a known gravity dual) we have a map
  between states on the cylinder, $R\times S^{d-1}$,
   and operators on the plane, $R^d$. The dimension of the operator  is equal to the
  energy of the corresponding state. (The scaling dimension
 tells us how the operator scales under the scaling transformation we mentioned after \nref{adsmep}.)
  This follows from the fact that we can go to a Euclidean cylinder. The Euclidean cylinder and
  the plane differ by an overall Weyl transformation of the metric
  $d(\log r)^2 + d\Omega^2 = { 1\over r^2}
[ dr^2 + r^2 d\Omega^2 ]$. Thus, they are equivalent in a CFT.
 An operator at the origin of the plane creates a state at fixed $r$ which can
be viewed as a state of the field theory on the cylinder, see figure \ref{planetocylinder}. This state-operator mapping is valid for any
conformal field theory.
AdS/CFT relates a state of the field theory on the cylinder with a state of the bulk theory in
global coordinates \nref{adsmet}.
 Since the symmetries on both sides are the same,
   we can divide the states, or operators,  according
 to their transformation laws under the conformal group. Such representations are characterized
 by the spin of the operator and its scaling dimension, $\Delta$.
   A simple example is the stress tensor operator, $T_{\mu \nu}$.
  This operator creates a graviton in $AdS$.
 The dimension of the stress tensor is $d$.
  Single trace operators are associated to single particle states in the bulk. Multitrace operators
  correspond to multiparticle states in the bulk.
  There exists a general argument, based on a simple analysis of Feynman diagrams, that shows that
  the dimensions of multitrace operators are the sum of the dimensions of each single trace component,
  up to $1/N^2$ corrections. In fact, the same analysis of Feynman diagrams shows that the large $N$
  limit of general gauge theories gives a string theory \cite{tHooft}. The argument does not specify precisely
  the kind of string theory we are supposed to get, it only says that we can organize the diagrams
  in terms of diagrams we can draw on a sphere plus the ones on the torus, etc. Each time we increase
  the genus of the surface, we get an additional power of $1/N^2$. This looks like a string theory
  with a string coupling $g_s \sim 1/N$. The strings we have in the bulk are precisely the strings that
  are suggested by this argument.

The preceeding argument says that the large $N$ limit is necessary to have a weakly coupled gravity
theory. This does not mean that we are restricted to the linearized solutions. In a weakly coupled
gravity theory we can consider full classical non-linear solutions of the equations, such as the
black hole solutions we discussed above. Weak coupling means that we can neglect quantum gravity
corrections or loop diagrams.

 In string theory the graviton is the lowest oscillation mode of a string. The gravitational
 coupling we discussed above is related to the interaction strength between strings.
 However,    we have another condition for the validity of the gravity approximation. In gravity
 we treat the graviton as a pointlike particle and we ignore all the massive string states.
  The typical size of the graviton is given by the string length, $l_s$, an additional parameter
  beyond the planck scale. In order to ignore the rest of the string states
  we need
  \be \label{gravap}
  { L_{AdS} \over l_s}  \gg 1 ~, ~~~~~~~~~{\rm for~gravity~to~be~a~good~approximation}
  \ee
   This condition is simply saying that the typical size of the space should
  be much bigger than the intrinsic size of the graviton in string theory.
  This condition is important because in many concrete examples, we have to make sure that this
  condition is met, otherwise gravity will give the wrong answers, even if we have a large number of
  fields!. If \nref{gravap} is not valid, then we should consider the full string theory in $AdS$.
  A salient feature of string theory is that there are  massive  string  states  of
  higher spin, $S>2$. In fact, in a large $N$ gauge theory we can easily write down single
  trace operators with higher spin, such as $Tr[ F_{\mu \nu}  D_+^S F^{\mu \nu } ] $, with
$D_+$ is a derivative along a null direction. Such operators
  have relatively small scaling dimensions at weak coupling, in this case $\Delta = 4 + S$.
  These give rise to particles of spin $S$ whose bulk
  mass is comparable to the inverse $AdS$ radius. Such light string states render the
  Einstein  gravity approximation invalid. Thus,
  {\it in order to trust the gravity approximation, the field theory should necessarily be strongly interacting}. This is a necessary, but not sufficient, condition.
  The coupling should be strong enough to give a large dimension to all the higher spin single trace
operators  of the
  theory. Such masses are set by the parameter \nref{gravap}.
  In concrete examples we find that this quantity, \nref{gravap}, is proportional to a positive power of  the effective
  't Hooft coupling of the theory $g^2_{YM} N$. Here $g_{YM}$ is the coupling constant of the
   gauge theory. The extra factor of $N$ comes in because color correlated particles can exchange
   $N$ gluons, which enhances their interactions at large $N$. By taking a large value of $g^2_{YM} N$
   it is possible to give a large dimension to the higher spin states. We are left with a light graviton, and other lower spin states. In this cases, we expect that their interactions are those of
   Einstein gravity.

 \section{Scalar field in $AdS$ }

 In order to be  more precise about the correspondence between states in $AdS$ and states
in the boundary   it is necessary to do the quantum mechanics of a particle in $AdS$.
 Equivalently, we quantize the corresponding field in $AdS$.
 In this subsection we consider a massive scalar field in $AdS$ with an action
 \be \label{actscal}
 S = \int d^{d+1}x   \sqrt{g} \left[  ( \nabla \phi )^2 + m^2 \phi^2 \right]
 \ee
 Let us compute the energy spectrum in $AdS$ in global coordinates,
 \nref{adsmet}.
 Let us only focus on the ground state. We expect that it should have zero angular momentum.
 So we make an ansatz for the wavefunction $\phi = e^{ - i \omega t } F(r)$, with the boundary
 condition that $F(r) \to 0$ at infinity.
 By setting $\nabla^2 \phi - m^2 \phi =0 $ we get an $\omega$ dependent
 equation for $F(r)$ with two boundary conditions,
 one at infinity and one at the origin. It is an eigenvalue problem which gives quantized
 frequencies.
 It is possible to check that
 \be \label{ground}
  \phi = e^{ - i \Delta t } { 1 \over ( 1 + r^2 )^{\Delta \over 2 } }
  \ee
  with
\be \label{dim}
 \Delta = { d \over 2 } + \sqrt{ { d^2 \over 4 } + ( m L)^2 }
 \ee
 is a solution of the equation of motion with the right boundary conditions.
  We identify this solution as the ground state, since it has no
 oscillations in the radial direction. The energy is $\omega = \Delta$.
   The energies of all other states differ from this
 by an integer, $\omega_n = \Delta +n$.
  This is due to the fact that we can get all the other states from the action of
 the conformal generators, thus, their energies are determined by the conformal algebra.

  We get a wavefunction localized near the center of $AdS$. In the large mass limit
  $mL \gg 1$, we see that it becomes sharply localized at $r=0$,
  as we expect for a classical particle in $AdS$.
  For $m L$ of order one,  the wavefunction is extended over a region of order one in $r$, which
  corresponds to proper distances of order the $AdS$ radius.
  A particle of zero mass, $m=0$, has an integer energy $\Delta = d$. The case of the graviton gives
  an equation which is similar to that of a massless field, and also leads to $\Delta =d$, as expected
  from the dimension of the stress tensor.

It seems from \nref{dim} that the dimensions of operators are bounded below by $d$, which is
the dimension of a marginal operator in the field theory. However, there are two effects that allow
us to go to lower dimensions. First, there are some allowed ``tachyons'' in $AdS$. Namely, it is possible
for a field to have $ - d^2/4\leq  (mL)^2 <0$. In other words, if a field is only slightly tachyonic,
it is allowed \cite{BF}. The reason that it does not lead to an instability is due to the boundary conditions.
These boundary conditions force the field to have some kinetic energy in the radial direction which
overwhelms the negative energy of the mass term. In fact, we can check from \nref{dim} that such
states have positive energies. A second fact, is that in the range $ -d^2/4 \leq (mL)^2 < 1-d^2/4$ we
can have a second quantization prescription \cite{KW}.
To understand that,   note that if we choose the other sign for the square root in
\nref{dim}, then \nref{ground} is another solution of the equation of motion.
  For tachyons, both  solutions  decay as $r\to \infty$. So
we can set boundary conditions that removes any of  these. The quantization leading to \nref{dim} corresponds to removing the solution that decays more slowly as $r\to \infty$.

Let us now do a different computation that will further elucidate the relation between bulk fields
 and
boundary operators.
It is convenient to go to Euclidean space and to choose the Poincare coordinates
\nref{adsmep}.
 We can  consider the problem of computing
the path integral of this scalar field theory with fixed boundary conditions at the  boundary.
The quantum gravity problem in $AdS$   contains such a
problem: we have to do this for all the fields of the theory, including the graviton.

Let us consider the classical, or semiclassical, contribution to this problem.
This is given by finding the classical solution that obeys the boundary condition and evaluating the
action for this solution. If we set zero boundary conditions, then the field is zero and the action
is zero. If we set non-zero boundary conditions, the classical action gives us something interesting.
Since we have translation symmetry along the boundary directions we go to Fourier space and
write
$\phi = e^{i k x} f(k,z)$.
The wave equation becomes
\be \label{weq}
{d^2 f \over dz^2} + (1-d ) { 1 \over z }{ d f \over dz} - \left( k^2 + {(m L)^2 \over z^2}\right) f = 0
\ee
Near the boundary, for small $z$, there are two independent solutions behaving as
 as $f = z^\Delta$, or $f\sim z^{d-\Delta}$.
  We will put a boundary condition on the
largest component of the solution. Since that component of the solution depends on $z$, we
put a boundary condition at $z= \epsilon$ and set the boundary condition to
\be \label{bcfa}
\phi(x,z)|_{z=\epsilon}  = \phi_0(x) \epsilon^{d-\Delta}
\ee
The solution of \nref{weq}  that decays
at $z \to \infty$ is
\be
 f(k,z) = e^{ i k z } z^{d/2} K_\nu( k z) ~,~~~~~~~~~ \nu = \sqrt{ { d^2 \over 4 } + (mL)^2 }
\ee
where $K$ is a BesselK function.
In order to obey the boundary conditions we set
\be
\phi(k,z) = \phi_0(k) \epsilon^{d - \Delta}  { f(k,z)  \over f(k,\epsilon) }
\ee
We now insert this into the action \nref{actscal}.
We can integrate by parts and use the equations of motion. The computation then reduces to
a boundary term. For each Fourier mode we get \cite{Freedman}
\bea
S &=&   \phi_0(-\vec k) \epsilon^{d - \Delta} { 1 \over \epsilon ^{d}}  z d_z\phi(k,z)|_{z=\epsilon} =
   \phi_0(-k)\phi_0(\vec k)  \epsilon^{d -  2 \Delta}  { zd_z f(k,z)  \over f(k,\epsilon) }
\cr
S &=&  \phi_0(-k)\phi_0(\vec k) \left[  \epsilon^{- 2 \nu }  {\rm Polynomial}[k^2 \epsilon^2 ]  -
 |k|^{2 \nu } 2^{ - 2 \nu} { \Gamma(-\nu) \over \Gamma(\nu) } 2 \nu  \right] \label{corrfi}
\eea
Note that the first term contains divergent terms when $\epsilon \to 0$. These terms are analytic
in momentum and, upon Fourier transformation, give terms that are local in position space.
These terms were to be expected since the boundary conditions we are considering are such that
the field grows towards the boundary.  From the field theory point of view these
 divergencies can be viewed as UV divergencies.
 On the other hand, the last  term in \nref{corrfi} gives a non-local
 contribution in position space and represents the interesting part of the correlator.
 Transformed back to position space this gives
 \be \label{acfi}
  S = - { 2 \nu \Gamma(\Delta) \over \pi^{d \over 2 } \Gamma(\nu) } L^{d-1}
  \int d^d x d^dy {  \phi_0(x) \phi_0(y) \over |x-y|^{ 2 \Delta } }
 \ee

 The AdS/CFT dictionary states that this computation with fixed boundary conditions is related
 to the generating function of correlation functions for the corresponding operator in the field theory
 \cite{GKP,wittenhol}.
 In other words, for a field $\phi$ related to the single trace
 operator ${\cal O}$ we have the equality
 \be
  Z_{\rm Gravity}[\phi_0(x) ] = Z_{\rm Field ~Theory} [\phi_0(x) ] = \langle e^{ \int d^d x \phi_0(x) {\cal O}(x) } \rangle
  \ee
  The leading approximation to the gravity answer is given by evaluating the classical action and it is
  given by  $e^{ - S} $, with $S$ in \nref{acfi}. Correlation functions of operators are then given
  by
  \be \label{gravitycorr}
   \langle {\cal O}(x_1) \cdots {\cal O}(x_n) \rangle = { \delta \over \delta \phi_0(x_1)  } \cdots
   { \delta \over \delta \phi_0(x_1)}   Z_{\rm Gravity}[ \phi_0(x) ]
 \ee
 In the quadratic approximation the gravity answer is given by \nref{acfi} and the correlation functions
 factorize into products of two point functions. We can include interactions in the bulk. For example,
 we can have a $\phi^3$ bulk interaction.
Then the  leading approximation is given by considering the classical, but non-linear solution
 with these boundary conditions and evaluating the corresponding action. This can be computed
 perturbatively by evaluating Feynman-Witten diagrams in the bulk \cite{wittenhol}.

For each single trace operator we have a corresponding field in the bulk with a certain
boundary condition.
Among these fields is the graviton, associated to   the stress tensor.
The generating function of correlation functions of the stress tensor is obtained by considering
the field theory on a general boundary geometry $g^b_{\mu \nu}(x)$. At the classical level we find
a solution of Einstein's equations, $R_{\mu \nu } \propto g_{\mu \nu}$,  with $g^b_{\mu \nu}$ as a
boundary condition.
We insert this in the action and obtain the quantity $Z_{\rm Gravity}[g^b_{\mu \nu}(x)] \sim e^{ -S_E[g_{cl}]}$.
We can also view this quantity as the Hartle-Hawking wavefunction of the universe in the Euclidean region.
As a first step, one can expand Einstein's equations to quadratic order and
  compute the two point function.  The action for each polarization component is similar to that
  of a massless scalar field. In this case,   the $\epsilon$ dependent
 factor in \nref{bcfa} drops out. So it makes sense to compute the absolute normalization
 of the two point function. This two point function of the stress tensor is another measure of
 the degrees of freedom of the theory. It is proportional to the overall coefficient in the Einstein
 action, which is the quantity $c$ introduced earlier, \nref{centralexpr}. In other words, we schematically have
 $T_{\mu \nu} (x) T_{\sigma \delta}(0)  = c { t_{\mu\nu\sigma\delta}  \over |x|^{2 d} } $ where
 $t_{\mu \nu \sigma \delta} $ is an $x$ dependent
  tensor taking into account the fact that the stress tensor is traceless and conserved. In fact,
  since the classical gravity action contains $c$ as an overall factor we conclude that
  to leading order in $1/c$, and in the gravity approximation,
  all stress tensor correlators are proportional to $c$. They  are universal for any field theory
  that has a gravity dual, for each spacetime dimension.
   Such correlators are not universal in quantum field theory
  (except in two dimensions, or $d=2$). The universality arises only in the gravity approximation and it
  is removed by higher derivative corrections to the action. Stringy corrections give rise to these
  higher derivative corrections.
  Similarly, the coefficient that appears in this computation is equal to the coefficient appearing
  in the computation of the thermal free energy (up to universal constants). Again, this does not hold for
  general field theories in $d>2$. It does hold in $d=2$.

Another interesting case  is a gauge field in $AdS$. This corresponds to a conserved current
on the boundary theory. Thus a gauge symmetry in the bulk corresponds to a global symmetry on the
boundary.  We have an exactly conserved charge in the boundary theory.
Due to black holes, the only way to ensure that we have a conserved charge is to have a gauge symmetry
in the bulk.

          \section{The ${\cal N}=4$ Super Yang Mills/$AdS_5 \times S^5$ example}

          The previous discussion was completely general. In order to be  specific, let
          us discuss one particular explicit example of a dual pair. We will first discuss the
          field theory, then the gravity theory. We will show how various objects match on
          the two sides.

          We  consider a four dimensional field theory that is similar to quantum chromodynamics.
          In quantum chromodynamics we have a gauge field $A_\mu$ which is a (traceless)
          $3\times 3$ matrix in the adjoint representation of $SU(3)$.  The action is
          \be
           S = - { 1 \over 4 g^2_{YM} } \int d^4 x Tr[ F_{\mu \nu} F^{\mu \nu} ] ~,~~~~~~~~~ F_{\mu \nu } = [ \partial_\mu + A_\mu , \partial_\nu + A_{\nu } ]
           \ee
           We can consider the generalization of this theory to a gauge group $SU(N)$, or $U(N)$, where
           $A_\mu$ is now an $N\times N$ matrix.
           In QCD we also have fermions which transform in the fundamental representation. Here
           we will do something different. We   add fermions that transform in the adjoint representation. The reason is that we would like to construct a theory that is supersymmetric.
          Supersymmetry is a powerful tool to check many of the predictions of the duality.
           The existence of the duality does not rely on supersymmetry, but is easier to find a
           dual pair when we have supersymmetry. Supersymmetry is a symmetry that relates bosons and
           fermions.
           In a supersymmetric theory the bosons and their  fermionic partners are in the same representation of
           the gauge group. If we simply add a majorana fermion in the adjoint we get an
           ${\cal N}=1$ supersymmetric theory. This theory is not conformal quantum
           mechanically, it has a beta function, as in the theory with no fermions.
           If instead we add four fermions, $\chi_\alpha$,  and six scalars, $\phi^I$,
            all in the adjoint, and with special couplings, we get a theory which has maximal supersymmetry, an ${\cal N}=4$ supersymmetric theory \cite{Green:1982sw}.
          The lagrangian of this theory is completely determined by supersymmetry and the choice of
           the gauge group. It has the schematic form
           \bea
           S & = & - { 1 \over 4 g^2_{YM} }
           \int d^4 x Tr\left[ F^2 + 2 (D_\mu \Phi^I )^2 + \chi    \not{   D} \chi + \chi [ \Phi , \chi] -
           \sum_{IJ} [\Phi^I,\Phi^J]^2 \right]  +
           \cr
           &&+  { \theta \over 8 \pi^2 }  \int Tr[ F \wedge F ]
           \eea
          We have two constants  which are the coupling constant, $g_{YM}^2$, and the  $\theta$
        angle. All relative coefficients in the lagrangian
          are determined by supersymmetry. The fields are all in a single supermultiplet under
          supersymmetry. This theory is classically and { \it quantum mechanically }
           conformal invariant.
          In other words, its beta function is zero. Thus, unlike QCD, it does not become more weakly
          coupled as you go to high energies. The coupling is set once and for all. If it is weak, it is
          weak at all energies, if it is strong it is strong at all energies.
          The effective coupling constant is
          \be
           \lambda = g^2_{YM} N
           \ee
           The extra factor of $N$ arises as follows.
          If we have two fields whose color and anticolor are  entangled, or summed over,
           then there are $N$ gluons that
          can be exchanged between them that preserve this entanglement.
          The theory has an $SO(6)$, or $SU(4)$, R-symmetry that rotates the six scalars into each other,
          and also rotates the fermions. An ``R'' symmetry is a symmetry that does not commute with
          supersymmetry. This is the case here because bosons and fermions are in different representations of $SU(4)$.

          Now let us discuss the gravity theory. It is a string theory, which gives rise to a
          quantum mechanically consistent gravity theory. Since we started from a supersymmetric
          gauge theory, we also expect to have a supersymmetric string theory. There are well known
          supersymmetric string theories in ten dimensions. In particular, there is one theory
          that contains only closed oriented strings called type IIB.
          This string theory reduces at long distances to a gravity theory. It is a
          supergravity theory called, not surprisingly, type IIB supergravity \cite{Schwarz:1983qr}.
           This is a theory
          that contains the metric, plus other massless fields  required
          by supersymmetry. In particular, this theory contains a five form field strength $F_{\mu_1 \cdots \mu_5}$, completely antisymmetric in the indices. It is also constrained to be self
          dual $F_5 = * F_5$.
           It is analogous to the two form field strength
          $F_{\mu \nu}$ of electromagnetism. In four dimensions we can have charged black hole solutions
          which involve the metric and the electric (or magnetic) two form field strength.
          In particular, the near horizon solution of an extremal black hole has the geometry
          $AdS_2 \times S^2$ with a two form flux  on the $AdS_2 $ (or the $S^2$) for an electrically
          (or a magnetically) charged black hole.
          Something similar arises in ten dimensions. There is a solution of the equations of the form
          $AdS_5 \times S^5$ with a five form along both the $AdS_5$ and $S^5$ directions. We have
           both electric and magnetic fields due to the self duality constraint on $F_5$.
          The Dirac quantization condition says that magnetic fluxes on an $S^2$ is quantized.  In the string theory case,   the flux on the $S^5$ is
           also quantized
          \be
           \int_{S^5}  F_5 \propto  N
           \ee
           This number is the same as the number of colors of the gauge theory.

           The equations of motion of ten dimensional supergravity that are relevant for us follow from the action
 \be
            S = { 1 \over (2 \pi)^7 l_p^{8} } \int d^{10} x \sqrt{g} ( R - F_5^2)
            \ee
            plus the self duality constraint, $ F_5 = * F_5$.
            The equations of motion relate the radius of $AdS_5$ and $S^5$ to $N$.
            In fact, we find that  both radii are given by
 ${L^4 \over \ell_p^4 } = 4 \pi N $.
           In string theory we also have the string length given by $l_s = g_s^{-1/4} l_p$. This sets
           the string tension, $T = { 1 \over 2 \pi l_s^2 }$. $g_s$ determines  the interaction strength between strings. It is given
         by the vacuum expectation value of one of the massless fields of the ten dimensional
         theory, $g_s = \langle e^{\phi} \rangle$.
          The gravity theory has another massless scalar field $\chi$. This second field is an
           axion, with a periodicity $\chi \to \chi + 2 \pi$.
           These two fields are associated
          to the two parameters $g^2_{YM}$ and $\theta$ that we had in the lagrangian.
          It is natural to identify $\theta $ with the expectation value, or boundary condition,
          for $\chi$ and $g^2_{YM} $ to the string coupling $g_s$, $g^2_{YM} = 4 \pi g_s$.
The precise numerical coefficient can be set by the physics of D-branes \cite{PolchinskiD}, or by
using the S-duality of both theories.
          After doing this, one can write the $AdS_5$ and $S^5$ radii in terms of the Yang Mills
          quantities
          \be \label{radiusexp}
           { L^4 \over l_s^4 } = 4 \pi g_s N = g^2_{YM} N = \lambda  ~,~~~~~~~~~~{L^4 \over l_p^4 }=  4 \pi N
           \ee
As we discussed in general, in order to have a weakly coupled bulk theory we need, $N\gg 1$.
 In addition, in order to trust the Einstein gravity approximation, we need a large effective coupling.
 Thus, we have the following situation:
\bea
 g^2_{YM} N&\gg& 1 :   ~~~{\rm Gravity~ is  ~good,~~~ gauge~theory~is~strongly ~coupled \notag
}\\
g^2_{YM} N&\ll& 1 :   ~~~{\rm Gravity~ is ~not ~good,~~~ gauge~theory~is~weakly ~coupled} \notag
\eea
In these two extreme regimes it is easy to do computations using one of the two descriptions.

The 't Hooft limit \cite{tHooft}, which gives planar diagrams, corresponds to
$N\to \infty$ with $g^2_{YM} N $ fixed.
It is sometimes useful to take the 't Hooft limit first and obtain a free string theory in
the bulk and then vary the 't Hooft coupling $\lambda$ from weak to strong, so that we change
the $AdS$ radius in string units. The string is governed by a two dimensional field theory whose
target space is $AdS$  (plus the $S^5$ and some fermionic dimensions). This two dimensional field
theory is weakly coupled if the $AdS$ radius is large and strongly coupled when the radius is small
or the gauge theory is weakly coupled. For
values of order one, $g^2_{YM} N \sim 1 $,  one needs to use the full  string theory description or
solve the full planar gauge theory.

 The ${\cal N}=4$ super Yang Mills theory has an S-duality symmetry   which exchanges weak and strong
 coupling.
 One is tempted to go to strong coupling and then use S-duality in
 order to get a weakly coupled theory again. This does not work.
  The bulk theory also has an S duality symmetry. These two S-duality symmetries are in one to one
  correspondence. So in order to test whether we can trust the gravity description, first we do
  S-duality on both sides to send $g_s<1$ and then we apply the criterium stated above.

It is interesting to return to  the problem of comparing the thermal free energy of the gauge
theory and the gravity theory. This time we keep track of the numerical coefficients. We consider
the field theory in $R^3 \times S^1_\beta$. The free energy at weak coupling is given by
the usual formula
\bea
- \beta F &=& V \int { d^3 k \over ( 2 \pi)^3 } \left[ n_{\rm bosons} \log { 1 \over( 1 - e^{ - \beta |\vec k|} ) }
+ n_{\rm fermions} \log (1 + e^{ - \beta |\vec k | } )\right]
\cr
- \beta F &=& { \pi^2 \over 6 } { V  } N^2 T^3 ~,~~~~~~~~~\beta = 1/T\label{freeweak}
\eea
where we used $n_{\rm bosons} = n_{\rm fermions} = 8 N^2$.
At strong coupling we consider the Euclidean
 black brane solution, with $\tau \sim \tau + \beta$,
\be
ds^2 = L^2 \left[   (1 - { z^4 \over z^4_0}  ) { d\tau^2 \over z^2 } + { d z^2 \over z^2 (1 - { z^4 \over z^4_0} ) } +
{ d x^2 \over z^2 } \right]
\ee
which is simply related to the large mass limit of \nref{adsbhmet}. We can relate $\beta = \pi z_0$
by demanding no singularity at $z=z_0$, as usual.
The entropy is given by the usual Bekenstein-Hawking formula \cite{Peet}
\bea
 S &=& { \rm Area \over 4 G_N} =  { L^8   V_{S^5} \over 4 G_{N,10} z_0^3  } = { \pi^2 \over 2 } { V   } N^2 T^3
 \eea
From the entropy we can simply compute the free energy. We get
\be
 - \beta F = S/4 = { \pi^2 \over 8 } { V   } N^2 T^3 \label{freestrong}
 \ee
We see that there is a factor of $3/4$ difference between \nref{freestrong} and \nref{freeweak}.
This does {\it not} represent a disagreement with AdS/CFT. On the contrary, it is a prediction
for how the free energy changes between weak and strong coupling.
Under general large $N$ arguments we expect the free energy to have the form
\be
{  F(\lambda,N) \over  F(\lambda=0,N) } = f_0(\lambda) + { 1 \over N^2 } f_1(\lambda) + \cdots
\ee
We expect that $f_0(\lambda)$ goes smoothly between $f_0=1$ at $\lambda =0$ and $f_0 = { 3 \over 4} $ at
$\lambda \gg 1 $. In fact, the leading corrections from both values has been computed and they go
in the naively expected direction \cite{GKT,Taylor} . In this example, the function $f_0$  constant at large $\lambda$.
There are examples where
 this function  goes as $f_0 \sim { 1 \over \sqrt{\lambda } }$ for large $\lambda$, \cite{ABJM,Drukker}.

If we are interested in computing the free energy of super Yang Mills at strong coupling, we
can do it by using the gravity result \nref{freestrong}.

The existence of the  $S^5$ is related to the $SO(6)$ symmetry of the theory.
The killing vectors generating the $S^5$ isometries give rise to gauge fields in $AdS_5$.
 These are the gauge fields associated to global symmetries that we expected in general.
 Of course, in other gauge/gravity duality examples,
  one can also have global symmetries which are not associated to a Kaluza Klein gauge
 field.

There are many  observables that have a simple geometric description at strong coupling.
In fact, the strings (and the branes) of string theory can end on the boundary and they correspond
to various types of operators in the boundary theory. For example, a Wilson loop operator
$Tr[P e^{ \oint_{\cal C}  A }] $ can be computed in terms of a string in the bulk that ends
on the boundary along the contour ${\cal C}$. At strong coupling, the leading approximation is given
just by the area of the surface that ends on this contour. At finite coupling, we need to do the worldsheet quantization of this theory. In other words, we need to sum over all surfaces that end on this contour.
Certain Wilson loops can be computed exactly using techniques that rely on supersymmetry, confirming
the predictions of the duality \cite{Pestun}.

The gauge theory contains scalar fields. The potential for these scalar fields have flat directions.
Namely, it is possible to give  expectation values to the fields in such a way that the vacuum
energy continues to be zero. This spontaneously breaks the conformal symmetry. At high energies the
conformal symmetry is restored, but it is broken at low energies. These flat directions corresponds
to expectation values for  the scalar fields which are diagonal matrices. As a simple example we can
set $\Phi^1 = {\rm diag} (a,0,\cdots ,0)$ and all the rest to zero. This breaks the gauge group
from $U(N) \to U(1)\times U(N-1)$. In the gravity dual, it corresponds to setting a D3 brane
at a position $z \sim 1/a$ in the Poincare coordinates \nref{adsmep}. One would
 expect that the gravitational potential pushes the brane towards the horizon. This force is
 precisely balanced by an electric repulsion which is provided by the presence of the electric five
 form field strength. The masssless fields living
on this D3 brane correspond to the fields in the $U(1)$ factor. The massive $W$ bosons arising from
the Higgs mechanism correspond to strings that
go from the brane to the horizon.
It is interesting that one can write  the solutions that correspond to general
vacuum expectation values.
We s write
\be
ds^2 = f^{-1/2} ( - dt^2 + d\vec x ^{\, 2} ) + f^{1/2} ( d\vec y^{\, 2} ) \label{bgen}
\ee
\be
  f = 4 \pi \sum_{i } { l_p^4 \over |\vec y - \vec y_ i|^4 } \nonumber
 \ee
 Here $\vec x$ is a three dimensional vector and $\vec y $ is a six dimensional vector.
 The $\vec y_i$ are related to the vacuum expectation values of the scalar fields, $\vec \Phi = {
 \rm diag} ( \vec y_1 , \vec y_2 , \cdots , \vec y_N ) $.
 This solution looks like a multicentered black brane. In principle we cannot trust the solution
 near a single center since the curvature is very large. However, in situations where we have many
 coincident centers, we can trust the solution. For example, if we break $U(2 N) \to U(N) \times U(N)$ by
giving the expectation value $\Phi^1 = {\rm diag}( a,\cdots,a, 0, \cdots, 0)$, with $N$ $a's$, we can
trust the solution everywhere. In the UV, for large $|\vec y|$, we have a single $AdS$ geometry which
splits into two $AdS$ throats with smaller radii as we go to lower values of $|\vec y|$. This describes the
corresponding flow in the gauge theory from the UV to the IR where we have two decoupled conformal
field theories. This is an example of a geometry which is only asymptotically $AdS$ near the boundary
but it is different in the interior.

\begin{figure}
\begin{center}
  \includegraphics[height=1.6in]{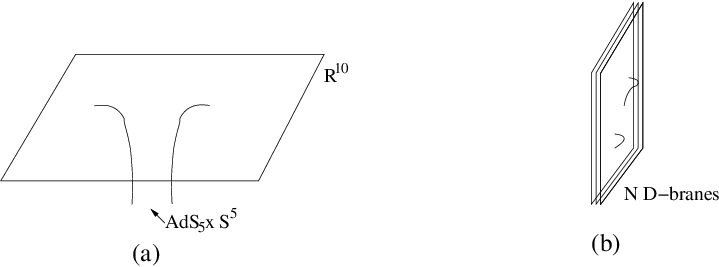}\\
  \caption{  (a) The geometry of the black 3-brane solution \nref{bgen} with \nref{newf}. Far away we
  have ten dimensional flat space. Near the horizon we have $AdS_5 \times S^5$.
  (b) The D-brane description. D-branes excitations are described by open strings living on them. They can start
  and end on any of $N$ D-branes so we have $N^2$ of them. At low energies they give rise to a $U(N)$ gauge theory: $ {\cal N}=4$ super Yang-Mills.
   }\label{DbranesAdS}
  \end{center}
\end{figure}

It is instructive to consider the solution \nref{bgen} with \cite{HS}
\be
\label{newf}
f = 1 + { 4 \pi N  l_p^4 \over |y|^4}
\ee
This enables us to give a physical derivation of the gauge-gravity duality for this example
\cite{JM}.
This solution goes to ten dimensional flat space for $|\vec y|\gg N^{1/4} l_p$.
It represents an extremal black three brane, see figure \ref{DbranesAdS}. It is extended along 1+3 of the spacetime dimensions, labelled by $t, \vec x$, and it is localized in six of them, labelled by $\vec y$.
The near horizon geometry of this black D3 brane is obtained by going to small values of $y$ and
dropping the $1$ in \nref{newf}.  When the string coupling is very small, $g_s N\ll1$, this system
can be descibed as a set of $N$ D3 branes. D3 branes are  solitonic defects that exist in
string theory \cite{JoeP}. They are described in terms of an extremely simple string theory construction.
This construction tells us that we get $N=4$ super Yang Mills at low energies. In fact, it is easy
to understand the scalar fields: they come from the motion of the branes in the six transverse dimensions.
We can view the gauge fields as arising from supersymmetry. A system of $N$ identical branes is expected to have an ordinary $S_N$ permutation symmetry. However, for these branes, this symmetry is
 enlarged into a
full $U(N)$ gauge symmetry.  Thus, we have two descriptions for the brane: first as a black brane and
second as a set of D-branes.  We can now take the low energy limit of each of these descriptions.
 The low energy limit of the
D-branes gives us the ${\cal N}=4$ $ U(N)$ super Yang Mills theory. The low energy limit on the
gravity side  corresponds to going very close to the horizon of the black three brane. There, the large
redshift factor ( the fact that $f^{-1/2} \to 0$) gives a very low energy to all the particles living in that
near horizon region. This region is simply $AdS_5 \times S^5$. Assuming that these two descriptions
are equivalent we get the gauge-gravity duality.

\section{The spectrum of states  or  operators}

In this case we can make a complete dictionary between the massless fields in the bulk and
operators in the field theory. The massless ten dimensional fields can be expanded in
spherical harmonics on the $S^5$. In addition, they  fill supermultiplets. It is interesting
to note that we have 32 supercharges in the bulk theory. With this large amount of supercharges
a generic supermultiplet would contain states with spins bigger than two. However, we can
have special BPS multiplets with spins only up to two. Thus all the massless particles of
the ten dimensional theory should be in special BPS multiplets. In ten flat dimensions this is
only possible if the particles are massless. In $AdS_5\times S^5$, this is only possible if
the $AdS$ energy is fixed in terms of the $SO(6)$ charge. In the field theory these are
in multiplets that contain the operators $Tr[ \Phi^{(I_1} \Phi^{I_2} \cdots \Phi^{I_J)} ]$ where
the $SO(6)$ indices are symmetrized and the traces extracted. These operators are in the
same representation as the spherical harmonics on  $S^5$ with angular momentum $J$.
Their dimension is $\Delta = J$, at all values of the coupling because it is a BPS state.
Here we see the power of supersymmetry allowing us to compute these dimensions for all values
of the coupling.
These operators
correspond to a special field in the bulk theory, which is a deformation of the $S^5$ and the 5-form
 field strength. The rest of the supergravity  fields are related to this one
by supersymmetry.

It is interesting to consider the fate of other operators.
As we mentioned above we can consider higher spin operators.
It is simpler to understand the mechanism that gives them a large dimension
by considering operators with large charges.
Let us consider $Z = \Phi^1 + i \Phi^2$, and the operator $Tr[Z^J]$.
If we now add some derivatives, such as $Tr[ D_+ Z ZZ D_+ ZZZ \cdots]$, then at
weak coupling the dimension of the operator is the same independently of the order.
As we turn on the coupling, the Hamiltonian, or the dilatation operator starts moving these
derivatives. In some sense, we can view the chain of $Z$s as defining a lattice. The fact that only
planar diagrams contribute implies that the interactions are short range on this lattice. The range
increases as we increase the order in perturbation theory. So the derivatives start moving around
and they gain a kinetic energy that depends on their ``momentum'' along the chain of $Z$'s. More
explicitly, the operators that diagonalize the Hamiltonian (of the dilatation operator) have the
schematic form
\be
{\cal O} \sim \sum_{l} e^{i p l } Tr[ D_+ Z Z^l D_+Z Z^{J-l-2} ]  + \cdots
\ee
We used the cyclicity of the trace to set the first derivative on the first spot. The dots represent
extra terms that can appear when $l$ or $J-l-2$ are small, and are important to precisely
quantize the momentum. The momentum $p$ is quantized with an expression of the form
 $p_n \sim  2 \pi n/J + o(1/J^2 )$ where the subleading term depends on the extra terms that appear when
the  two derivatives ``cross'' each other.
A derivative with zero momentum, is a derivative which can be pulled out of the trace and acts as an
ordinary derivative. This is just an element of the conformal group and it does not give rise to spin, but
to ordinary orbital angular momentum in $AdS$. Thus, in order to get spin, we need derivatives which
have some momentum, and thus some kinetic energy. This kinetic energy increases as $\lambda$ increases.
Thus, for large $\lambda$, all the states that have non-zero momentum get a large energy. This is
specially true for a short string (small $J$), where the momentum has to be relatively large, due to the momentum
quantization condition.
This was just a qualitative argument. The exact computation of these energies requires considerable
technology and it employs a deep ``integrability'' symmetry of the planar gauge theory, or the
corresponding string theory \cite{Beisert:2010jr}. For the lightest spin four state (Konishi
multiplet),
these energies have been computed for any $\lambda$ in \cite{Gromov:2009zb}. They behave as expected and
are in complete agreement with $AdS/CFT$.
Namely, they go from an order one value at weak coupling to the strong coupling answer which is
$\Delta = 2 \lambda^{1/4} $. This strong coupling answer is computed as follows. When the AdS radius
is large in string units the massive string states feel almost as if they were in flat space. The
lightest massive string state in flat space has mass $m^2 = 4/l_s^2$ \cite{Polchinski:1998rq}. Using \nref{dim} and \nref{radiusexp} this gives $\Delta \sim L m \sim 2 \lambda^{1/4}$.

\section{The radial direction}

\begin{figure}
\begin{center}
  \includegraphics[height=2.in]{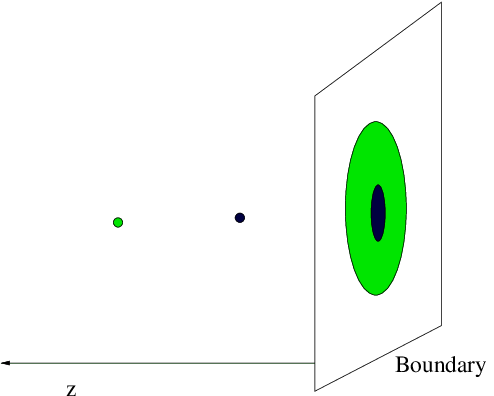}\\
  \caption{ The size/radius correspondence. In the CFT we have excitations which have a size. We see the same object with two different sizes, related by a dilatation. They correspond to two particles
   with the same proper size in $AdS$ but located at different values of the radial position of
   $AdS$.  }\label{radial}
  \end{center}
\end{figure}

One of the crucial elements of the gauge gravity duality is the emergence of an extra
``radial'' dimension, the $z$ coordinate in \nref{adsmep}.
Let us discuss this in  more detail and in some generality.
In ordinary physics we are used to particles that are typically massive. These particles have
a quantum state described by the three spatial positions. Even if they have internal constituents, like
the proton or an atom (ignoring spin),
we can describe the state by simply giving its spatial momentum or position.
 Of course, its energy is
determined by its mass.
In a scale invariant theory we cannot have massive particles. Naively we would say that all the particles are
massless. On the other hand these massless particles interact in a non-trivial way and they cannot be viewed
 as good asymptotic states. This is true even in large $N$ gauge theories. However, in large
$N$ gauge theories we have some weakly interacting excitations. They are the objects that are created by
acting with single trace operators on the vacuum. These objects are characterized by the 4-momentum of the
operator. Notice that we have one more component of the momentum, as compared to an ordinary massive particle. For a simple operator, such as $Tr[F_{\mu \nu} F^{\mu \nu}]$ the state created at zero coupling with a given
four momentum is a pair of gluons which sum up to the total four
 momentum. As we increase the coupling we start
producing more and more gluons via a showering process.
For these CFT states we
 can specify arbitrarily the value of $k^2 = k_\mu k^\mu$. In the CFT we can view these as objects
 that have a position and, in addition, a size. The size is one
  more continuous variable that we need to specify the characterize the state. This is the reason that
  we need to give four continuous quantum numbers to specify a state in the four dimensional CFT.
 These states are not particles in the CFT,  they are particles in $AdS$.
 We can say that the size of the state in the CFT is related to the position along the radial direction in
  $AdS$, see figure \ref{radial}.
In the coordinates \nref{adsmep}, the size is proportional to $z$.
In fact, particles in $AdS$ are a simple way to
parametrize the representations of the conformal group. In other words, unitary representations of the conformal
group correspond, in a one to one mapping, to particles, or fields, in $AdS$ together with a boundary condition.
 This is a completely general
mathematical result. We saw this explicitly above for the case of a scalar field. A representation is characterized
by the value of the scaling  dimension, which in turn determines the mass of the field.

Now, if this is so general, why don't all theories have gravity duals?. Well, to some extent we can say that they
all do. However, the gravity dual could be a strongly coupled theory in the bulk.
 Large $N$ theories give
weakly coupled string duals. However, they can be highly stringy. We need some additional conditions that ensure
an approximate locality of the interactions in the bulk. In particular we need locality within an
 $AdS$ radius. A necessary condition is that all the higher spin fields
have large anomalous dimensions. It might be that this is a sufficient condition, but this has not been clearly
demonstrated from the axioms of  conformal field theories.  Though, or course, this is expected to hold
if we assume bulk locality.

\begin{figure}
\begin{center}
  \includegraphics[height=1.5in]{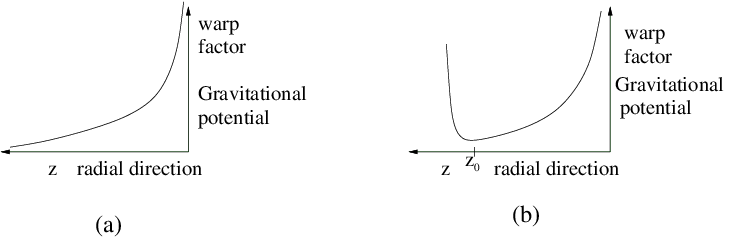}\\
  \caption{(a) The behavior of the warp factor, or gravitational potential in the $AdS$ case. It rises
   to infinity at the boundary and it goes to zero towards the interior. A particle is pushed towards
   ever larger values of $z$. In the field theory this corresponds to an excitation expanding in size.
   (b) Warp factor in a theory with a mass gap. The warp factor has a minimum
 and excitations minimize their energy by sitting at $z_0$. Typically the space ends, in a smooth way, at
$z=z_0$.
 In the dual field theory excitations have a preferred size, like the size of a proton in QCD.
   }\label{warpfactor}
  \end{center}
\end{figure}

Even though we focused on conformal field theories,
the gauge gravity duality is also valid for non-conformal theories \cite{IMSY,wittenthermal,KS}.
 In those cases the metric has the form
$ds^2 = w(z)^2 ( dx^2 + dz^2)$. As in the conformal case, $w$ rises rapidly when we approach
the boundary.
In these cases the size is also related to the $z$ direction. However, since we do not have a precise scaling
symmetry, the physical behavior of boundary objects of different sizes is different. The same happens in the bulk, particles at different positions in the $z$ direction see a geometry with different properties.
In some cases the geometric description fails for small $z$ or large $z$. In particular, this happens
when the gauge theory coupling becomes weak in the UV or the IR \cite{IMSY}. In such examples, the gravity
description is a good approximation only for distances scales (or energy scales) such that the gauge
theory coupling is large.
One can consider quantum field theories with a mass gap. The corresponding gravity configurations are
such that
  the warp factor has a minimum value at some position $z_0$.  In several examples one   finds that the space ends at $z_0$ because some
 other dimensions (similar to the $S^5$, above) are shrinking smoothly at $z = z_0$.
 A massive
particle minimizes its energy by sitting at $z_0$. Of course, its wavefunction is concentrated
 around $z_0$, see \ref{warpfactor}. The precise shape of the warp factor has to be computed by solving
 the bulk equations.

\begin{thebibliography}{99}


\bibitem{JM}
  J.~M.~Maldacena,
  Adv.\ Theor.\ Math.\ Phys.\  {\bf 2}, 231-252 (1998).
  [hep-th/9711200].

\bibitem{GKP}
  S.~S.~Gubser, I.~R.~Klebanov, A.~M.~Polyakov,
  Phys.\ Lett.\  {\bf B428}, 105-114 (1998).
  [hep-th/9802109].

\bibitem{wittenhol}
  E.~Witten,
  Adv.\ Theor.\ Math.\ Phys.\  {\bf 2}, 253-291 (1998).
  [hep-th/9802150].

\bibitem{tHooft}
  G.~'t Hooft,
  Nucl.\ Phys.\  {\bf B72}, 461 (1974).

\bibitem{BF}
  P.~Breitenlohner, D.~Z.~Freedman,
  Phys.\ Lett.\  {\bf B115}, 197 (1982).


\bibitem{KW}
  I.~R.~Klebanov, E.~Witten,
  Nucl.\ Phys.\  {\bf B556}, 89-114 (1999).
  [hep-th/9905104].

\bibitem{Freedman}
  D.~Z.~Freedman, S.~D.~Mathur, A.~Matusis, L.~Rastelli,
  Nucl.\ Phys.\  {\bf B546}, 96-118 (1999).
  [hep-th/9804058].

\bibitem{Green:1982sw}
  M.~B.~Green, J.~H.~Schwarz, L.~Brink,
  Nucl.\ Phys.\  {\bf B198}, 474-492 (1982).


\bibitem{Schwarz:1983qr}
  J.~H.~Schwarz,
  Nucl.\ Phys.\  {\bf B226}, 269 (1983).

\bibitem{PolchinskiD}
  J.~Polchinski, S.~Chaudhuri, C.~V.~Johnson,
  [hep-th/9602052].

\bibitem{Peet}
  S.~S.~Gubser, I.~R.~Klebanov, A.~W.~Peet,
  Phys.\ Rev.\  {\bf D54}, 3915-3919 (1996).
  [hep-th/9602135].



\bibitem{GKT}
  S.~S.~Gubser, I.~R.~Klebanov, A.~A.~Tseytlin,
  Nucl.\ Phys.\  {\bf B534}, 202-222 (1998).
  [hep-th/9805156].

\bibitem{Taylor}
  A.~Fotopoulos, T.~R.~Taylor,
  Phys.\ Rev.\  {\bf D59}, 061701 (1999).
  [hep-th/9811224].


\bibitem{ABJM}
  O.~Aharony, O.~Bergman, D.~L.~Jafferis, J.~Maldacena,
  JHEP {\bf 0810}, 091 (2008).
  [arXiv:0806.1218 [hep-th]].

\bibitem{Drukker}
  N.~Drukker, M.~Marino, P.~Putrov,
 [arXiv:1007.3837 [hep-th]].



\bibitem{Pestun}
  V.~Pestun,
  [arXiv:0712.2824 [hep-th]].


\bibitem{HS}
  G.~T.~Horowitz, A.~Strominger,
  Nucl.\ Phys.\  {\bf B360}, 197-209 (1991).


\bibitem{JoeP}
  J.~Polchinski,
  Phys.\ Rev.\ Lett.\  {\bf 75}, 4724-4727 (1995).
  [hep-th/9510017].


\bibitem{Beisert:2010jr}
  N.~Beisert  {\it et al.},
  ``Review of AdS/CFT Integrability: An Overview,''
  [arXiv:1012.3982 [hep-th]].




\bibitem{Gromov:2009zb}
  N.~Gromov, V.~Kazakov, P.~Vieira,
  Phys.\ Rev.\ Lett.\  {\bf 104}, 211601 (2010).
  [arXiv:0906.4240 [hep-th]].




\bibitem{Polchinski:1998rq}
  J.~Polchinski,
  Cambridge, UK: Univ. Pr. (1998) 402 p.


\bibitem{IMSY}
  N.~Itzhaki, J.~M.~Maldacena, J.~Sonnenschein, S.~Yankielowicz,
  Phys.\ Rev.\  {\bf D58}, 046004 (1998).
  [hep-th/9802042].


\bibitem{wittenthermal}
  E.~Witten,
  Adv.\ Theor.\ Math.\ Phys.\  {\bf 2}, 505-532 (1998).
  [hep-th/9803131].



\bibitem{KS}
  I.~R.~Klebanov, M.~J.~Strassler,
  JHEP {\bf 0008}, 052 (2000).
  [arXiv:hep-th/0007191 [hep-th]].



\end{thebibliography}
\end{document}